\documentclass[12pt]{article}
\usepackage{epsf}
\epsfverbosetrue
\catcode`@=11
\def\citer{\@ifnextchar
[{\@tempswatrue\@citexr}{\@tempswafalse\@citexr[]}}

\def\@citexr[#1]#2{\if@filesw\immediate\write\@auxout{\string\citation{#2}}\fi
  \def\@citea{}\@cite{\@for\@citeb:=#2\do
    {\@citea\def\@citea{--\penalty\@m}\@ifundefined
       {b@\@citeb}{{\bf ?}\@warning
       {Citation `\@citeb' on page \thepage \space undefined}}%
\hbox{\csname b@\@citeb\endcsname}}}{#1}}
\catcode`@=12
\newcommand{\beq}{\begin{equation}}
\newcommand{\eeq}{\end{equation}}
\newcommand{\beqa}{\begin{eqnarray}}
\newcommand{\eeqa}{\end{eqnarray}}

\def\simgt{\rlap{\lower 3.5 pt \hbox{$\mathchar \sim$}} \raise 1pt \hbox {$>$}}
\def\simlt{\rlap{\lower 3.5 pt \hbox{$\mathchar \sim$}} \raise 1pt \hbox {$<$}}

\def \MSbar {\vbox{\hrule\kern 1pt\hbox{\rm MS}}}

\def\lsim{\simlt}

\def\eq#1{Eq.~(\ref{#1})}

\def\junk#1{}
\begin{document}

\begin{titlepage}

\begin{flushright}
   Edinburgh 2000/02
\\ hep-ph/0003035
\end{flushright}

\vspace*{1cm}

\begin{center} {\Large Treatment of Heavy Quarks in Deeply
 	Inelastic Scattering}

\vspace*{1cm}

{\sc Michael~Kr\"amer$^{1}$}, {\sc Fredrick~I.~Olness$^2$},
and
{\sc Davison~E.~Soper$^{3}$}

\vspace*{1cm}

${}^1${\it Department of Physics and Astronomy, University of
Edinburgh, Edinburgh EH9 3JZ, Scotland}\\[1mm]
${}^2${\it Department of Physics, Southern Methodist University,
	Dallas, Texas 75275, USA}\\[1mm]
${}^3${\it Institute of Theoretical Science, University of Oregon,
	Eugene OR 97403, USA}

\end{center}

\vspace*{1cm}

\begin{abstract}
We investigate a simplified version of the ACOT prescription for
calculating deeply inelastic scattering from  $Q^2$ values near the
squared mass $M_H^2$ of a heavy quark to $Q^2$ much larger than $M_H^2$.
\end{abstract}

\end{titlepage}

\section{Introduction}

The inclusion of heavy quark effects in deeply inelastic scattering
(\mbox{DIS}) is an interesting theoretical problem involving two hard
scales in a perturbative analysis. This issue is also of
phenomenological importance. The charm contribution to the total
structure function $F_2$ at small $x$, at HERA, is sizeable, up to
$25\%$. Through the charm contribution to scaling violations, the
treatment of charm also has a significant impact on the interpretation
of fixed target deeply inelastic scattering data. Thus, a proper
description of charm contributions to deeply inelastic scattering is
required for a global analysis of structure function data and a
precise extraction of the parton densities in the proton.

At scales $Q\;\simlt\; M_H$, the contribution to deeply inelastic
scattering of a heavy quark of mass $M_H$ can be calculated in the
so-called fixed-flavor-number (FFN) prescription from hard processes
initiated by light quarks ($u,d,s,\dots$) and gluons, where all
effects of the heavy quark ($H$) are contained in the perturbative
coefficient functions. This prescription incorporates the correct
threshold behavior, but for large scales, $Q\gg M_H$, the coefficient
functions at higher orders in $\alpha_s$ contain potentially large
logarithms $\ln^i(Q^2/M_H^2)$, which may need to be summed
\cite{Barnett:1988jw,LRSN,or}. Such a summation can be achieved by
including the heavy quark as an active parton in the proton.  The
simplest approach incorporating this idea is the so-called zero-mass
variable-flavor-number (ZM-VFN) prescription, where heavy quarks are
omitted entirely below some scale $Q_0 \approx M_H$ and included as
massless partons above this threshold. This prescription has been used
in global analyses of parton distributions for many years, but it has
an error of order $M_H^2/Q^2$ and is not suited for quantitative
analyses unless $Q\gg M_H$.

Considerable effort has been devoted to including heavy quark
effects in deeply inelastic scattering in such a way that the
calculated structure functions match those of the FFN prescription
in the region $Q \approx M_H$ while they match those of the ZM-VFN
prescription for $Q \gg M_H$. Two prescriptions of this sort, the
Aivazis--Collins--Olness--Tung (ACOT) \cite{ACOT} and the
Thorne--Roberts \cite{TR} prescriptions have been used in recent
global analyses of parton distributions \cite{LaiTun97a,MRST}.  More
recently, additional variable-flavor-number prescriptions with
non-zero mass have been defined in the literature \cite{buza}. If one
could sum perturbation theory, the calculated structure functions
should be identical for any prescription that does not neglect the
heavy quark mass. However, the way of ordering the perturbative
expansion is not unique, so that the results generally differ at any
finite order in perturbation theory.

In this paper, we will investigate a modification of the ACOT
prescription advocated by Collins~\cite{collins}. It has
the advantage of being simple to state and of allowing relatively
simple calculations. This simplicity should be convenient for
phenomenological analyses at the Born level. In addition, it could be
crucial for implementing a variable-flavor-number prescription with
non-zero mass at next-to-leading order in global analyses of parton
distributions.

\section{Schemes and partons}

We consider the structure function $F_2(x,Q)$. The observable
structure function can be written in terms of parton distribution
functions $f$ and a calculable partonic structure function $\hat F_2$
as
\begin{equation}
F_2(x,Q) =
\int_x^1\! d\xi\, \sum_a f_{a/p}(\xi,\mu) \
\widehat F_2(a,x/\xi,\mu/Q,\alpha_s(\mu))
+{\cal O}(\Lambda^2/Q^2).
\label{factor}
\end{equation}
This factorization formula has corrections of order $\Lambda^2/Q^2$,
where $\Lambda$ is a typical scale of hadronic physics, perhaps the
mass of the rho meson. Here $\mu$ is the
renormalization/fac\-to\-ri\-za\-tion scale. (For simplicity, we do
not distinguish these two scales.) The partonic $\hat F_2$ has a
perturbative expansion in powers of $\alpha_s(\mu)$. There is a sum
over parton types $a = g,u,\bar u, d,
\bar d, \dots$, and there is a function $f_{a/p}$ for each parton.

To begin, we must pick a scheme for the definition of $\alpha_s$ and
for the definition of the parton distribution functions. In fact, we
will use multiple schemes, following the prescription of
\cite{collins}. Each scheme is designated by the number $N$ of
``active'' quark flavors. Scheme $N$ is designed to be most useful for
physical scales $Q$ in the range $M_N \,\lsim\, Q\, \lsim\, M_{N+1}$,
where $M_N$ is the mass of the $N$th quark flavor.

We define what we mean by the running coupling $\alpha_s^N(\mu)$ in
scheme $N$ by defining how we perform renormalization. We renormalize
using the CWZ prescription \cite{CWZ}.  Briefly, divergences involving
active parton loops are removed with an \MSbar\ subtraction, but when
active parton external lines couple to a loop containing
``non-partonic'' lines (quarks $N+1, N+2, \dots$) the renormalization
is by subtraction at zero external momentum.  The running of
$\alpha_s^N(\mu)$ is controlled by the usual \MSbar\
renormalization group equation with the contributions from quarks
$1,\dots,N$ in the beta function.

In the scheme $N$ there are parton distributions for gluons and for
quark flavors $1,\dots,N$, but not for quark flavors $N+1,\dots$. The
parton distribution functions are defined to be proton matrix elements
of certain operators \cite{partondef}. The operator products are
ultraviolet divergent and are renormalized according to the CWZ
prescription. They obey a renormalization group equation -- the usual
DGLAP equation -- in which contributions from quark flavors $N+1,\dots$
do not appear in the kernel.

Note that the parton distribution functions are non-perturbative
objects. There is no question of neglecting any masses in the
definition. The quark masses do not appear in the kernel of the
evolution equation, but this is because renormalization counter terms
are mass independent, not because the parton distributions themselves
are mass independent.

Note also that by using CWZ renormalization throughout, quark loops for
the non-partonic flavors decouple from calculations when the momentum
scale is small compared to $M_{N+1}$. Thus for $Q \ll M_{N+1}$ it is a
good approximation to leave the non-partonic flavors out of calculations
altogether. (But leaving heavy flavors out of calculations is not part
of the definition of scheme $N$ as used here; it is a separate
approximation.) When one does leave the non-partonic flavors out of
calculations, the remaining renormalizations are via the
\MSbar\ prescription. Thus one commonly refers to parton distributions
in the prescription described here as \MSbar\ parton distributions.

There is a perturbative connection between schemes $N$ and $N+1$
which we can represent by:
\begin{equation}
f_{a/p}^{N+1} (x,\mu) =
f_{a/p}^{N} (x,\mu) +
\sum_{b} \int_x^1 { d\xi \over \xi}
\ A_{ab}(x/\xi,\mu/M_{N+1}, \alpha_s(\mu)) \
f_{b/p}^{N} (\xi,\mu).
\label{pdfmatching}
\end{equation}
Here the index $a$ runs over $g$, the $N$ light quarks and their
anti-quarks, and the heavy quark $H$ and anti-quark $\bar H$. For $a =
H,\bar H$ we define $f_{a/p}^{N} (x,\mu) = 0$ in the first term on the
right hand side. The index $b$ runs over only the gluon and light
quarks. At order $\alpha_s$, the only nonvanishing $A_{Hb}$ is
$A_{Hg}$.  Thus we may say that the $H$ distribution arises
perturbatively from $g \to H\bar H$.

The $\alpha_s^1$ terms in the perturbative expansion of the $A_{ab}$
vanish at $\mu = M_{N+1}$. Thus one derives the simple matching
conditon \cite{CollinsTung}: $f_{a/p}^{N+1} (x,M_{N+1}) = f_{a/p}^{N}
(x,M_{N+1}) + {\cal O}(\alpha_s^2)$. The analogous connection is known
at order $\alpha_s^2$ but we do not repeat it here.\footnote{%
At order $\alpha_s^2$, the matching at $\mu = M_{N+1}$ is no longer
continous in general \cite{buza}.}

\section{Why is there an ambiguity?}

When we construct the hard scattering cross section in the presence of
heavy quarks, there is a factorization ambiguity that is not present in
the case of light quarks. Let us see why. Consider two enormously
simplified examples that illustrate the principle.

First, suppose that there are only gluons and one light quark $L$ with
mass $M_L \lsim \Lambda$. Suppose, in addition, that the light quark
is its own anti-particle, so that a quark $L$ and the anti-quark $\bar
L$ are the same particle. Both the gluon and the light quark are
considered to be active partons. In order to simplify the notation,
let us take a moment $\int dx\, x^n F_2(x,Q)$ of $F_2$, so that we get
a factorization formula involving the corresponding moments of
$\widehat F_2$ and of the parton distributions. The dependence on the
moment number will not be indicated. To further simplify the notation,
let us set the factorization and renormalization scale $\mu$ to
$Q$. Then
\begin{equation}
F_2(Q) \sim
\widehat F_2(L,\alpha_s(Q))\,f_{L/p}(Q)
+\widehat F_2(g,\alpha_s(Q))\,f_{g/p}(Q).
\label{whyL}
\end{equation}
Now $F_2$ is an observable, so its definition is fixed. We have
defined the parton distribution functions, and the two parton
distribution functions are independent. Thus this equation all but
fixes the definition of $\widehat F_2(L,\alpha_s(Q))$ and $\widehat
F_2(g,\alpha_s(Q))$. The only possible modification would be to add
terms proportional to powers of $M_L^2/Q^2$ to $\widehat
F_2(a,\alpha_s(Q))$, at the cost of subtracting the same terms from
the power suppressed remainder, ${\cal O}(\Lambda^2/Q^2)$, in the
factorization formula (\eq{factor}). The simplest solution, which is
uniformly adopted, is not to allow a $M_L^2/Q^2$ dependence in
$\widehat F_2(a,\alpha_s(Q))$.

Now suppose that there are only gluons and one (self-conjugate) heavy
quark $H$ with mass $M_H \gg \Lambda$. Then in the scheme in which
both the gluon and the heavy quark are considered to be active partons
we have
\begin{equation}
F_2(Q) \sim
\widehat F_2(H,M_H/Q,\alpha_s(Q))\,f_{H/p}(Q)
+\widehat F_2(g,M_H/Q,\alpha_s(Q))\,f_{g/p}(Q).
\label{whyH}
\end{equation}
Here the $\widehat F_2(a,M_H/Q,\alpha_s(Q))$ (a=g,H) depend on the
heavy quark mass and we cannot move terms of order $M_H^2/Q^2$ into
the power suppressed corrections because only terms of order
$\Lambda^2/Q^2$ are allowed there. Thus it seems that we have no
freedom. But we do: $f_{H/p}(Q)$ and $f_{g/p}(Q)$ are not
independent. Since, according to \eq{pdfmatching}, heavy quarks evolve
from gluons, we have a relation of the form
\begin{equation}
f_{H/p}(Q) =
V_{H/g}(\ln(Q/M_H),\alpha_s(Q))\,f_{g/p}(Q).
\label{whyevolve}
\end{equation}
Here $V$ has a perturbative expansion that is obtained by solving the
evolution equation. The first term has the form $V \sim \alpha_s \gamma
\ln(Q/M)$ where $\gamma$ is a constant. Using \eq{whyevolve} in 
\eq{whyH} we obtain
\begin{equation}
F_2(Q) \sim
\left\{
\widehat F_2(H,M_H/Q,\alpha_s(Q))\,V_{H/g}(\ln(Q/M_H),\alpha_s(Q))
+\widehat F_2(g,M_H/Q,\alpha_s(Q))
\right\}
f_{g/p}(Q).
\label{whyH2}
\end{equation}

Evidently, there is some freedom to move pieces from the first term in
braces to the second. There is a constraint. For $M_H/Q \to 0$, it is
possible to neglect $M_H$ in the calculation of the $\widehat F_2$.
On the other hand, $V(\ln(Q/M_H),\alpha_s(Q))$ does not have a smooth
$M_H/Q \to 0$ limit. This is not a problem in applications because by
solving the evolution equation one sums the leading logarithms
$[\alpha_s \ln(Q/M_H)]^n$ in $V$. It is important not to undo this
summation. Thus the factorization scheme that we adopt should have the
property that the functions $\widehat F_2(a,M_H/Q,\alpha_s(Q))$ have a
finite limit as $M_H/Q \to 0$. This still leaves us the option of
adding a term like $c \times (M_H^2/Q^2)$ to $\widehat
F_2(H,M_H/Q,\alpha_s(Q))$ and subtracting $c \times (M_H^2/Q^2)\,V$
from $\widehat F_2(g,M_H/Q,\alpha_s(Q))$.

Suppose now that we have calculated $\widehat
F_2(g,M_H/Q,\alpha_s(Q))$ and $\widehat F_2(H,M_H/Q,\alpha_s(Q))$ in
some convenient prescription -- for example the prescription analyzed
in \cite{collins} based on ``on-shell'' heavy quarks.  Then we could
define a new prescription with
\begin{equation}
\widehat F_2^{\rm new}(H,M_H/Q,\alpha_s(Q))
=
\widehat F_2^{\rm old}(H,0,\alpha_s(Q))
\label{transformH}
\end{equation}
and
\begin{eqnarray}
&&
\widehat F_2^{\rm new}(g,M_H/Q,\alpha_s(Q))
=
\widehat F_2^{\rm old}(g,M_H/Q,\alpha_s(Q))
\nonumber\\
&&\qquad +
\left\{
\widehat F_2^{\rm old}(H,M_H/Q,\alpha_s(Q))
-
\widehat F_2^{\rm old}(H,0,\alpha_s(Q))
\right\}
\times
\,V_{H/g}(\ln(Q/M_H),\alpha_s(Q)) .
\label{transformg}
\end{eqnarray}
In the following subsection, we shall give a prescription \cite{collins}
based on this observation.

\section{A prescription for resolving the ambiguity}

Having seen the main idea, let us put the parton indices and the
momentum fraction variables back. Suppose that there are $N+1$ quark
flavors that we consider to be active. Let the heaviest active quark,
quark $N+1$, be labelled $H$. Suppose that quark $H$ has a mass that
is large compared to the hadronic mass scale $\Lambda$. In the $N+1$
flavor scheme the factorization equation is
\begin{equation}
F_2(x,Q) \sim
\int_x^1\! d\xi\, \sum_{a} f_{a/p}(\xi,\mu) \
\widehat F_2(a,x/\xi,M_H/Q,\mu/Q,\alpha_s(\mu)),
\label{whyfull1}
\end{equation}
where the sum over $a$ includes $a = H$ and $a = \bar H$.

The parton distributions for the gluon and $N+1$ quark flavors are
obtained from the distributions in the scheme with only $N$ quark
flavors. First we use the perturbative matching relation
(\eq{pdfmatching}) at a scale near $\mu = M_H$, then we use the
evolution equations to give the distribution functions at scale $\mu$.
Since there are more output functions than input, we obtain a
perturbative relation giving the heavy quark distribution functions at
scale $\mu$ in terms of the light quark and gluon distribution
functions at the same scale. This relation has the form
\begin{equation}
f_{H/p}(\xi,\mu) =
\int_\xi^1 {d\tau\over \tau} \sum_{a \ne H \bar H}
f_{a/p}(\tau,\mu)
\ V_{H/a}(\xi/\tau \, ;\mu/M_H,\alpha_s(\mu)),
\label{whyfull2}
\end{equation}
with an analogous equation for the heavy anti-quark $\bar H$. Inserting
this relation into \eq{whyfull1}, we have
\begin{equation}
F_2(x,Q) \sim
\int_x^1\! d\xi \sum_{a \ne H,\bar H} f_{a/p}(\xi,\mu) \
T_{a}(x/\xi,M_H/Q,\mu/Q,\alpha_s(\mu)),
\label{whyfull3}
\end{equation}
where
\begin{eqnarray}
&& T_{a}(z,M_H/Q,\mu/Q,\alpha_s(\mu))
=
\widehat F_2(a,z,M_H/Q,\mu/Q,\alpha_s(\mu))
\nonumber \\
&& \qquad +
\int_z^1 d\lambda  \
\widehat F_2(H,z/\lambda,M_H/Q,\mu/Q,\alpha_s(\mu))\
V_{H/a}(\lambda,\mu/M_H,\alpha_s(\mu))
\nonumber\\
&& \qquad +
\int_z^1 d\lambda  \
\widehat F_2(\bar H,z/\lambda,M_H/Q,\mu/Q,\alpha_s(\mu))\
V_{\bar H/a}(\lambda,\mu/M_H,\alpha_s(\mu)).
\label{whyfull4}
\end{eqnarray}
We see that we have the same situation as in the simple example given
earlier. Taking $\mu$ to be of order $Q$, one can shift contributions
of order $M_H^2/Q^2$ between the hard scattering functions $\widehat
F_2$ for $H$ and $\bar H$ and the corresponding functions for the
light quarks and the gluon while keeping the functions $T$ unchanged
for each light quark and gluon flavor $a$ and without ruining the
property that all of the functions $\widehat F_2$ have finite $M_H /Q
\to 0$ limits.

This freedom can be exploited to make the calculation of the functions
$\hat F_2$ simpler. In particular, we can adopt a prescription
proposed by Collins \cite{collins}:

\begin{quote}
{\it Simplified ACOT (S-ACOT) prescription}.
Set $M_H$ to zero in the calculation of the hard
scattering functions $\widehat F_2$ for incoming heavy quarks.
\end{quote}

This observation tremendously simplifies the calculation of $\widehat
F_2$ for $a = H$ as it reduces to that of the light-quark result.  For
example, at order $\alpha_s^1$, the $\widehat F_2$ functions for heavy
quarks and light quarks are independent of $M_H$, and the calculation
reduces to that of the simple massless result.  The $\alpha_s^1$ gluon
contribution to $\widehat F_2$ acquires an $M_H$ dependence when the
gluon couples to a heavy quark loop, which is probed by the virtual
photon ($g \gamma \to H \bar H $).

Note that the hard scattering functions $\widehat F_2$ obey a
renormalization group equation that is different from the standard
renormalization group equation that applies in the case of light
flavors only. To see why this is so, imagine that the $\widehat F_2$
functions obeyed the usual renormalization equation and that $\widehat
F_2(H,z/\lambda,M_H/Q,\mu/Q,\alpha_s(\mu))$ were independent of $M_H$
at some fixed value of $\mu$. Then $\widehat F_2$ for $a = H, \bar H$
would depend on $M_H$ at other values of $\mu$ because the standard
renormalization group equation mixes the heavy and light flavors and
would mix $M_H$ dependence into $\widehat
F_2(H,z/\lambda,M_H/Q,\mu/Q,\alpha_s(\mu))$.

This Simplified ACOT prescription has the advantage of being simple to
state. In addition, its calculational simplicity could be crucial for
applying variable-flavor-number prescriptions with mass in analyses
that go beyond first order in $\alpha_s$.

\section{The ACOT and S-ACOT prescriptions at first order}

In this section, we analyze the ACOT prescription and its simplified
version, the S-ACOT prescription, at order $\alpha_s^1$. We consider
neutral current deeply inelastic scattering from a proton target in a
$Q^2$ regime in which quarks $1, \dots, N$ can be considered as light
while a single quark, $H$, is considered to be heavy or light
depending on the value of $Q^2$. In order to keep the analysis simple,
contributions from all heavier quarks are ignored. We further simplify
the problem by supposing that the vector boson current that probes the
proton couples only to the heavy quark $H$ and its anti-quark $\bar
H$, but not to the lighter quarks or gluons. (In a crude
approximation, this is like taking ordinary deeply inelastic
scattering but demanding that $H$ appear in the final state.) Let $F$
be one of the structure functions $F_1$, $F_2 /x$ or $F_3$ for our
special vector boson. We choose the factorization and renormalization
scales $\mu$ equal to $Q$. We write the factorization formula for $F$
in the shorthand notation
\begin{equation}
F = \sum_a \widehat F_a \otimes f_{a/p}
+{\cal O}(\Lambda^2/Q^2),
\label{factorii}
\end{equation}
where the sum runs over all partons $a$ including $H$ and $\bar H$. The
$\otimes$ denotes a convolution, so that \eq{factorii} means
\begin{equation}
F(x,Q) = \sum_a \int {d\xi \over \xi}\
\widehat F_a(x/\xi, Q; \alpha_s(Q))\
f_{a/p}(\xi,\mu)\
+{\cal O}(\Lambda^2/Q^2).
\end{equation}
The functions $\widehat F_a$ in \eq{factorii} are the hard scattering
functions. They have an expansion in powers of $\alpha_s$.  Keeping
the first two terms in this expansion gives
\begin{eqnarray}
F &=&
  \widehat F^{(0)}_H \otimes f_{H/p}
+ \widehat F^{(0)}_{\bar H} \otimes f_{\bar H/p}
\nonumber\\
&& +\
  \widehat F^{(1)}_g \otimes f_{g/p}
+ \widehat F^{(1)}_H \otimes f_{H/p}
+ \widehat F^{(1)}_{\bar H} \otimes f_{\bar H/p}
\nonumber\\
&& +\
 {\cal O}(\alpha_s^2)
+{\cal O}(\Lambda^2/Q^2),
\label{master}
\end{eqnarray}
where $\widehat F^{(n)}_a$ is the order $\alpha_s^n$ contribution to
$\widehat F_a$.

In the ACOT prescription, we choose the order zero hard scattering
function for a heavy quark to be
\begin{equation}
\widehat F^{(0)}_{H,{\rm ACOT}} = F^{(0)}_H(M_H),
\label{ACOTH0}
\end{equation}
where $F^{(0)}_H(M_H)$ is the calculated structure function for
scattering from an initial state heavy quark of mass $M_H$ that is on
its mass-shell. (To be precise, the heavy quark transverse momentum is
taken to be zero, and the structure functions $F_1, F_2/x, F_3$ are
extracted from the tensor $W^{\mu\nu}$ using the usual formula but
with the quark momentum $k^\mu$ replaced by a light-like vector
$\tilde k^\mu = k^\mu - [M_H^2/(2\, u\cdot k)]\, u^\mu$, where $u^\mu$
is a light-like reference vector in the plane of $q^\mu$ and $k^\mu$.)
The function $F^{(0)}_H(M_H)$ is rather complicated (see
Ref.~\cite{ACOT,kretzer,schmidt}), so we do not reproduce it here. The
definition of $\widehat F^{(0)}_{\bar H,{\rm ACOT}}$ for a heavy
anti-quark is analogous.

The order $\alpha_s$ gluon hard scattering function in \eq{master} has
three pieces:
\begin{equation}
\widehat F^{(1)}_{g,{\rm ACOT}} =
  F^{(1)}_g(M_H)
- F^{(0)}_H(M_H) \otimes f^{(1)}_{H/g}(M_H)
- F^{(0)}_{\bar H}(M_H) \otimes f^{(1)}_{\bar H/g}(M_H).
\label{ACOTg1}
\end{equation}
Here $F^{(1)}_g(M_H)$ is the calculated structure function for
scattering from an initial state massless gluon that is on its
mass-shell, using graphs with a heavy quark loop and a heavy
anti-quark loop. The function $f^{(1)}_{H/g}(M_H)$ is the calculated
$\alpha_s^1$ contribution to the distribution of heavy quarks in an
on-mass-shell gluon. Similarly, $f^{(1)}_{\bar H/g}(M_H)$ is the
calculated $\alpha_s^1$ contribution to the distribution of heavy
anti-quarks in an on-mass-shell gluon. Both functions are given by
\begin{equation}
f_{H/g}^{(1)}(\xi,Q) = f_{\bar H/g}^{(1)}(\xi,Q)
=
\frac {\alpha_{s}(Q)}{2\pi} \
\ln \frac{Q^{2}}{M_H^{2}} \ P_{qg}^{(1)}(\xi),
\label{f1Hg}
\end{equation}
where $P_{q/g}$ is the usual {\it gluon} $\to $ {\it quark} splitting
function $P_{qg}(\xi) = T_F\,(\xi^2+ (1-\xi)^2)$.

The first order hard scattering function for a heavy quark has a
structure similar to that of the corresponding function for a gluon,
\begin{equation}
\widehat F^{(1)}_{H,{\rm ACOT}} =
  F^{(1)}_H(M_H)
- F^{(0)}_H(M_H) \otimes f^{(1)}_{H/H}(M_H).
\label{ACOTH1}
\end{equation}
Here $F^{(1)}_H(M_H)$ is the order $\alpha_s^1$ contribution to the
structure function for scattering from an initial state heavy quark
that is on its mass-shell, as given in \cite{kretzer}. The function
$f^{(1)}_{H/H}(M_H)$ is the calculated $\alpha_s^1$ contribution to
the distribution of heavy quarks in an on-mass-shell heavy quark,
\begin{equation}
f_{H/H}^{(1)}(\xi,Q)
=
C_F \frac {\alpha_{s}(Q)}{2\pi}
\left[
\frac{1+\xi^2}{1-\xi}\,
\left\{
\ln\!\left( \frac{Q^{2}}{(1- \xi)^2 M_H^{2}}\right)
-1
\right\}
 \right]_+ ,
\end{equation}
where the + subscript denotes the usual prescription,
\begin{equation}
\int_0^1 d\xi \ f(\xi)\, \Bigl[F(\xi)\Bigr]_+
= \int_0^1 d\xi \ \{f(\xi) - f(1)\}\,F(\xi).
\end{equation}
This result can be calculated easily from the definition
\cite{partondef} of \MSbar\ parton distribution functions or it can be
extracted from the ACOT subtraction terms in \cite{kretzer}.

This defines the ACOT prescription at order $\alpha_s^1$. The
prescription has two important properties.

\begin{itemize}
\item
{\it Property 1.} For $Q/M\to \infty$, the hard scattering functions in
\eq{master} approach the hard scattering functions of the ZM-VFN
prescription, in which the heavy quark $H$ is taken as a parton with
zero mass.
\end{itemize}
\noindent
To be specific, we have the relations:
\begin{eqnarray}
\widehat F^{(0)}_{H,\mbox{\scriptsize ZM-VFN}} &=& F^{(0)}_H(0),
\nonumber\\
\widehat F^{(1)}_{g,\mbox{\scriptsize ZM-VFN}} &=&
  F^{(1)}_g(0)
- F^{(0)}_H(0) \otimes f^{(1)}_{H/g}(0)
- F^{(0)}_{\bar H}(0) \otimes f^{(1)}_{\bar H/g}(0),
\nonumber\\
\widehat F^{(1)}_{H,\mbox{\scriptsize ZM-VFN}} &=&
  F^{(1)}_H(0)
- F^{(0)}_H(0) \otimes f^{(1)}_{H/H}(0).
\label{ZMscheme}
\end{eqnarray}
We should note that in a calculation with $M_H = 0$, there are
infrared divergences. Thus the calculations are performed in
$4-2\epsilon$ dimensions of space-time and $1/\epsilon$ poles appear.
The pole terms cancel in \eq{ZMscheme}. Instead of using dimensional
regulation and taking $\epsilon \to 0$, one could use an infrared
regulator mass $m$ and let $m \to 0$.

\begin{itemize}
\item
{\it Property 2.} For $Q$ of order $M_H$, the structure function is
that of the fixed-flavor-number prescription, up to corrections of
order
$\alpha_s^2$.
\end{itemize}
\noindent
In the fixed-flavor-number prescription we have
\begin{equation}
F = F^{(1)}_g(M_H) \otimes f_{g/p} +
 {\cal O}(\alpha_s^2)
+{\cal O}(\Lambda^2/Q^2).
\label{masterFFN}
\end{equation}
Property 2 follows because when the factorization scale, $\mu = Q$, is
of order $M$, the heavy quark distribution function is given by the
perturbative formula
\begin{equation}
f_{H/p} = f^{(1)}_{H/g} \otimes f_{g/p} + {\cal O}(\alpha_s^2).
\label{Hinglue}
\end{equation}
First of all, since $f_{H/p}$ is of order $\alpha_s$, the terms
\begin{equation}
\widehat F^{(1)}_H \otimes f_{H/p}
+ \widehat F^{(1)}_{\bar H} \otimes f_{\bar H/p}
\end{equation}
in \eq{master} are of order $\alpha_s^2$ and can be dropped.
This leaves
\begin{eqnarray}
 F &=&
 F^{(0)}_H(M_H) \otimes f_{H/p}
+ F^{(0)}_{\bar H}(M_H) \otimes f_{\bar H/p}
\nonumber\\
&+& \
  [F^{(1)}_g(M_H)
- F^{(0)}_H(M_H) \otimes f^{(1)}_{H/g}(M_H)
- F^{(0)}_{\bar H}(M_H) \otimes f^{(1)}_{\bar H/g}(M_H)]
   \otimes f_{g/p}
\nonumber\\
&+& \
 {\cal O}(\alpha_s^2)
+{\cal O}(\Lambda^2/Q^2).
\label{masterencore}
\end{eqnarray}
Inserting \eq{Hinglue} into \eq{masterencore}, we obtain
\eq{masterFFN}.

We can summarize properties 1 and 2 by saying that the ACOT
prescription interpolates between the zero-mass prescription for $M_H
\ll Q$ and the fixed-flavor-number prescription for $M_H \sim Q$.

What about the simplified ACOT prescription? We use the same formulas
as for the ACOT prescription, but set $M_H = 0$ in the hard scattering
functions for heavy quarks:
\begin{eqnarray}
\widehat F^{(0)}_{H,\mbox{\scriptsize S-ACOT}} &=& F^{(0)}_H(0),
\nonumber\\
\widehat F^{(1)}_{g,\mbox{\scriptsize S-ACOT}} &=&
  F^{(1)}_g(M_H)
- F^{(0)}_H(0) \otimes f^{(1)}_{H/g}(M_H)
- F^{(0)}_{\bar H}(0) \otimes f^{(1)}_{\bar H/g}(M_H),
\nonumber\\
\widehat F^{(1)}_{H,\mbox{\scriptsize S-ACOT}} &=&
  F^{(1)}_H(0)
- F^{(0)}_H(0) \otimes f^{(1)}_{H/H}(0).
\label{SACOTscheme}
\end{eqnarray}
Repeating the derivation just given, we see that properties 1 and 2
hold for the S-ACOT prescription. That is, the S-ACOT prescription also
interpolates between the zero-mass prescription for $M_H \ll Q$ and the
fixed-flavor-number prescription for $M_H \sim Q$. However the S-ACOT
prescription has the advantage that the functions $\widehat F_H$ are
easier to calculate.


\section{Comparison of different prescriptions \label{sec:compare}}

In this section, we investigate how well the matching properties among
the different prescriptions work. In the plots presented, the masses
and couplings have been fixed to be consistent with the values used in
the CTEQ4L/M fits, i.e. $m_c = 1.6$~ GeV, $m_b = 5$~ GeV
\cite{CTEQ}. Below the bottom threshold ${\rm n_f}=4$ active flavors
are used for $\alpha_s$, and the scale has been chosen as $\mu=Q$.

\subsection{Parton distribution matching}

In the threshold region, the structure function calculated in the
S-ACOT prescription matches that calculated in the FFN prescription,
in which the heavy quark does not appear as a parton. This matching,
{\it Property 2}, results from the fact that the heavy quark
distribution function $f_{H/p}(x,\mu)$, matches the approximate
function
\begin{equation}
\widetilde{f}_{H/p}(\mu)
=   f^{(1)}_{H/g}(\mu) \otimes f_{g/p}(\mu) .
\label{pdfmatch}
\end{equation}
Both functions vanish at $\mu = M_H$ and, for $\mu>M_H$, the difference
between them is of order $\alpha_s^2$. In this subsection, we study the
quality of this matching.

In Figs.~1--3 we plot $f_{H/p}(x,\mu)$ and
$\widetilde{f}_{H/p}(x,\mu)$ for the case of the charm quark. In
Fig.~1, we use CTEQ4L parton distributions, which are based on lowest
order evolution.  Fig.~1 reveals that the matching works very well
when the order $\alpha_s^1$ evolution kernel is used for
$f_{H/p}(x,\mu)$ and the order $\alpha_s^1$ perturbative expression
(\eq{pdfmatch}) is used for $\widetilde{f}$. In Fig.~2, we use CTEQ4M
parton distributions, which are based on NLO evolution. We observe a
mismatch because the full NLO evolution kernel is used for $f_{H/p}$
while the order $\alpha_s^1$ perturbative expansion is used for
$\widetilde{f}$. The difference between $f$ and $\widetilde{f}$ is of
order $\alpha_s^2$, but this difference is numerically quite
large. Finally, we see in Fig.~3 that a close match is restored when
the NLO evolution kernel is used for $f$ while $\tilde f$ is defined
using a modified version of \eq{pdfmatch} in which we replace the
leading order $g \to H$ evolution kernel in \eq{f1Hg} by the
next-to-leading order $g \to H$ evolution kernel.\footnote{For the
NLO perturbative expansion, we have included the $\alpha_s^2$
splitting kernels, $P^{(2)}_{j/i}(\xi)$.  For simplicity, we have
ignored iterated terms such as $P^{(1)}_{k/j} \otimes P^{(1)}_{j/i}$
which contribute as $\ln^2(\mu/M_H)$, and hence only play a role away
from threshold.}

We can draw two conclusions. First, the threshold matching discussed
in the previous section will work order by order for the perturbation
expansion.\footnote{ At order $\alpha_s^2$, {\it Property~2} is still
preserved even though the matching conditions on the parton
distribution functions are modified. The $\alpha_s^2$ matching
conditions shift the evolved parton distribution functions $f_{H/p}$,
but they also shift the ``perturbative" parton distribution functions
$\widetilde{f}_{H/p}$ leaving the difference unchanged up to order
$\alpha_s^3$. } Second, the order $\alpha_s^2$ terms in the evolution
kernel are quite large, so that the leading order calculations
illustrated in this paper may not be sufficient for obtaining accurate
predictions.\footnote{ One might ask whether retaining the heavy quark
mass in the kinematics (in the spirit of the ``slow-rescaling"
correction) might prove beneficial. The answer is that as long as the
order of the parton distribution functions and the hard scattering
coefficients are matched, this simply amounts to a shuffling of
$M_H/Q$ terms between the quark and gluon initiated terms, {\it cf.},
\eq{whyH2}.}

\def\dir{}  

\begin{figure}
\begin{center}
 \leavevmode
 \epsfxsize=0.50\hsize \epsfbox{\dir 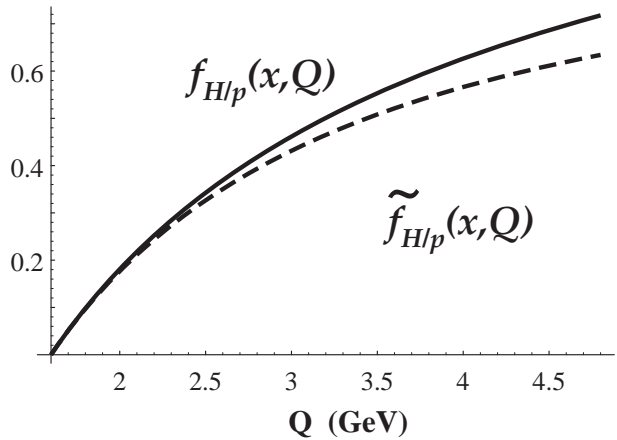}
\caption{ CTEQ4L charm quark density $f_{H/p}(x,Q)$ and
the approximate form $\tilde f_{H/p}(x,Q)$,
\eq{pdfmatch}, at $x=0.05$ as function of $Q$.
Since \eq{pdfmatch} is an order $\alpha_s^1$ perturbative
approximation and the evolution kernel for the CTEQ4L parton
distribution is also order $\alpha_s^1$, these curves match closely
at threshold.  }
\end{center}
 \label{fig:pdfi}
\end{figure}

\begin{figure}
\begin{center}
 \leavevmode
 \epsfxsize=0.50\hsize \epsfbox{\dir 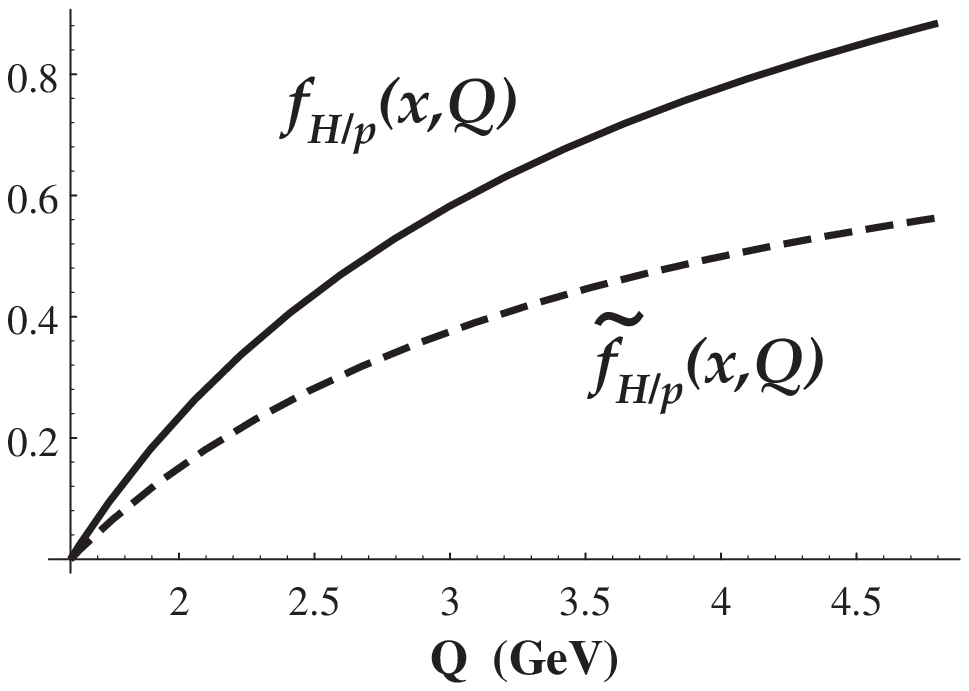}
\caption{ CTEQ4M charm quark density $f_{H/p}(x,Q)$ and
the approximate form $\tilde f_{H/p}(x,Q)$,
\eq{pdfmatch}, at $x=0.05$ as function of $Q$.
Since \eq{pdfmatch} is an order $\alpha_s^1$ perturbative
approximation while the evolution kernel for the CTEQ4M parton
distribution includes order $\alpha_s^2$ terms, these curves do
not match closely at threshold.  }
\end{center}
 \label{fig:pdfii}
\end{figure}

\begin{figure}
\begin{center}
 \leavevmode
 \epsfxsize=0.50\hsize \epsfbox{\dir 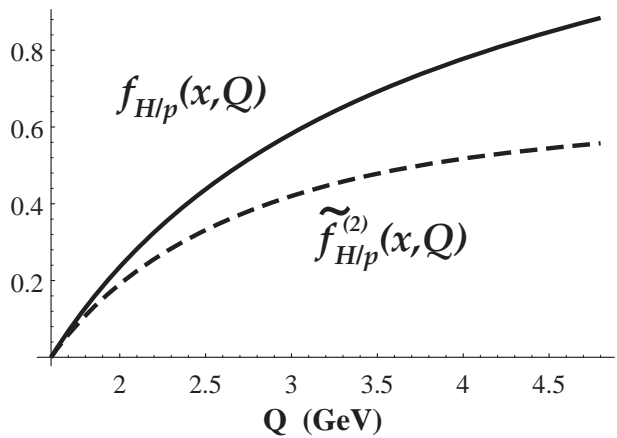}
\caption{ CTEQ4M charm quark density $f_{H/p}(x,Q)$ and an
approximate form $\tilde f_{H/p}^{(2)}(x,Q)$ at $x=0.05$ as
function of $Q$. The approximate form is based on an analogue of
\eq{pdfmatch} in which appropriate order $\alpha_s^2$ terms
are added. Since the calculation of $\tilde f_{H/p}^{(2)}(x,Q)$
uses the second order evolution kernel used for the CTEQ4M parton
distributions,  these curves match closely at threshold.  }
\end{center}
 \label{fig:pdfiii}
\end{figure}


\subsection{Structure function matching}

In this subsection, we examine predictions for $F_2^c(x,Q)$, which we
define here to be the contribution to $F_2(x,Q)$ from graphs in which
the current couples to a charm quark. We compare $F_2^c(x,Q)$
calculated with the S-ACOT prescription at order $\alpha_s$ with that
calculated with the original ACOT prescription, the ZM-VFN
prescription in which the charm quark can appear as a parton but has
zero mass, and the FFN prescription in which the charm quark has its
proper mass but does not appear as a parton. For simplicity, we take
$\mu=Q$. In our calculations, we evaluate the hard scattering
coefficients $\widehat F$ at order $\alpha_s$. Thus for the ZM-VFN,
ACOT, and S-ACOT prescriptions, the terms displayed in
\eq{master} are included. The functions $\widehat F$ are given
by \eq{ZMscheme} for ZM-VFN, by
Eqs.~(\ref{ACOTH0},\ref{ACOTg1},\ref{ACOTH1}) for ACOT, and by
\eq{SACOTscheme} for S-ACOT. For the FFN prescription, there is
only one term at order $\alpha_s$, as displayed in
\eq{masterFFN}.

In Fig.~4 we show $F_2^c(x,Q)$ as a function of $Q$ for $x = 0.1$.
Then in Fig.~5 we show $F_2^c(x,Q)$ as a function of $Q$ for $x =
0.001$.  In each case we display results using both the CTEQ4L and
CTEQ4M parton distributions.

When we use the CTEQ4L parton distributions, we notice that there is a
close match between the S-ACOT result and the FFN result near $Q =
m_c$.  Based on the results of the previous subsection, we expect this
matching to be degraded when we use CTEQ4M parton distributions
because of the important role played by the order $\alpha_s^2$ term in
the evolution kernel that is not matched in the lowest order
calculation of the hard scattering function. This degradation is seen
in the figures.

In the asymptotic regime, $Q \gg M$, we find the S-ACOT result
approximates the ZM-VFN result, as expected.

We observe that the ACOT and S-ACOT prescriptions are effectively
identical throughout the kinematic range. There is a slight difference
in the threshold region, but this is small in comparison to the size
of the $\mu$-variation (not shown). Hence the difference between the
ACOT and S-ACOT results is of no physical consequence. The fact that
the ACOT and S-ACOT prescriptions match extremely well throughout the
full kinematic range provides explicit numerical verification that the
S-ACOT prescription fully contains the physics.


\begin{figure}
\begin{center}
 \leavevmode
 \epsfxsize=0.47\hsize \epsfbox{\dir 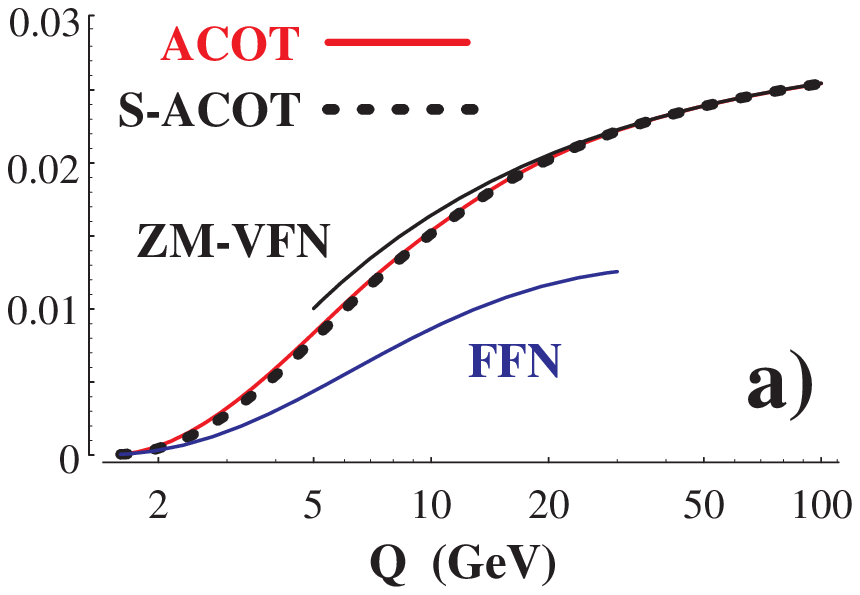} \hfill
 \epsfxsize=0.47\hsize \epsfbox{\dir 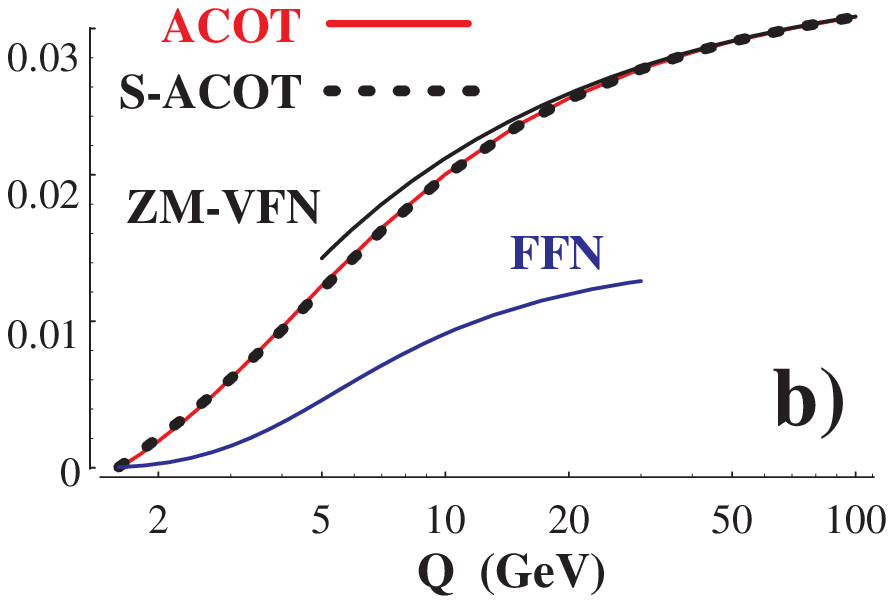}
\caption{
 $F_2^c$ for $x=0.1$ as a function of $Q$ as calculated using the
 ZM-VFN, FFN, ACOT, and S-ACOT prescriptions. The hard scattering
 coefficients are calculated to order $\alpha_s^1$.
 Plot~a) uses CTEQ4L parton densities and
 Plot~b) uses CTEQ4M parton densities.
}
\end{center}
\label{fig:charmi}
\end{figure}

\begin{figure}
\begin{center}
 \leavevmode
 \epsfxsize=0.47\hsize \epsfbox{\dir 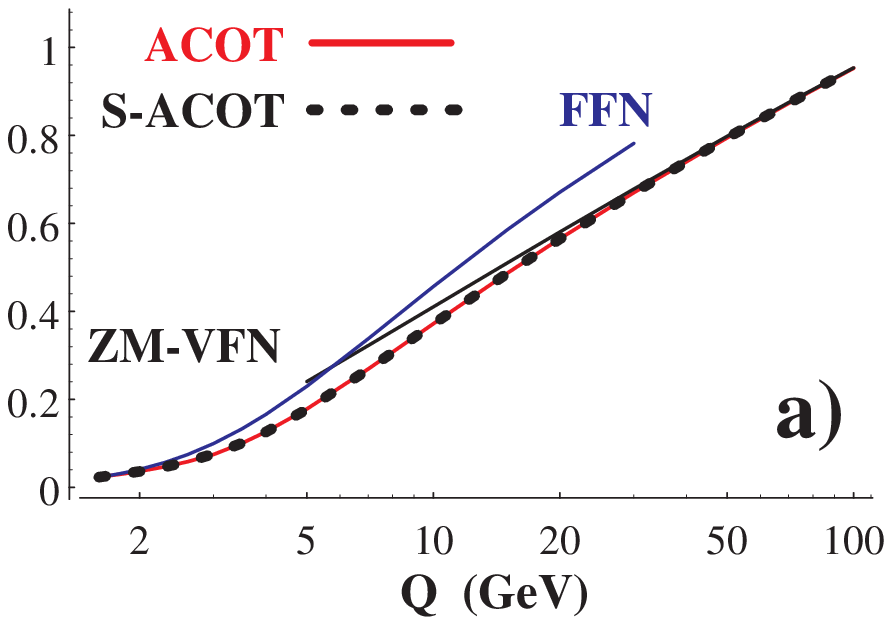} \hfill
 \epsfxsize=0.47\hsize \epsfbox{\dir 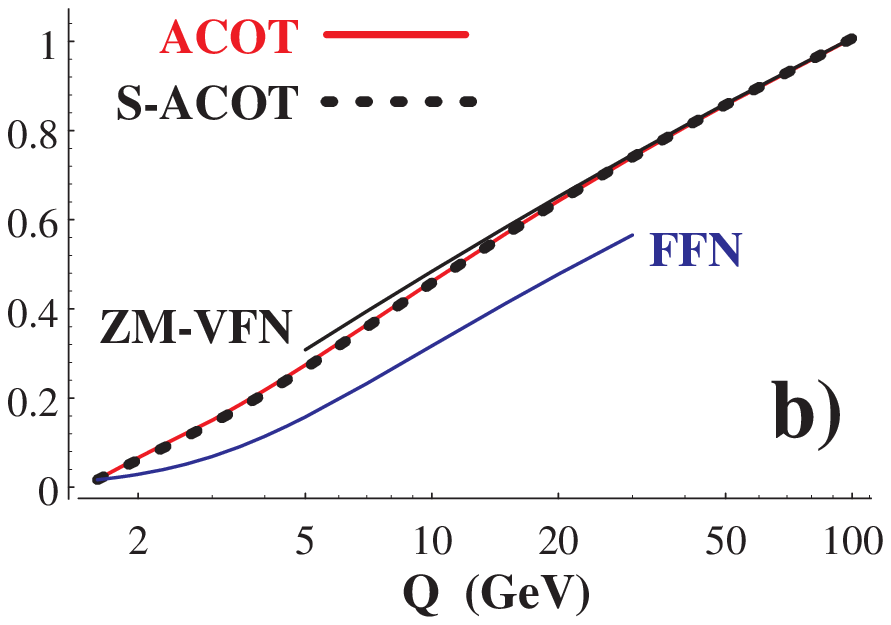}
\caption{
 $F_2^c$ for $x=0.001$ as a function of $Q$ as calculated using the
 ZM-VFN, FFN, ACOT, and S-ACOT prescriptions. The hard scattering
 coefficients are calculated to order $\alpha_s^1$.
 Plot~a) uses CTEQ4L parton densities and
 Plot~b) uses CTEQ4M parton densities.
}
\end{center}
\label{fig:charmii}
\end{figure}


\section{Conclusions and outlook}

We have performed a numerical study of different prescriptions for
dealing with the quark mass in heavy quark leptoproduction. We have
seen that the simplest prescription \cite{collins}, S-ACOT, is
numerically equivalent to the earlier ACOT prescription \cite{ACOT}.

The S-ACOT prescription is extensible order by order in $\alpha_s$.
At ${\cal O}(\alpha_s^1)$ we already find a significant simplification
in the S-ACOT prescription as compared with the ACOT prescription. We
expect that this simplification becomes even more dramatic at higher
orders.

We have not attempted to implement the S-ACOT prescription at order
$\alpha_s^2$, but we note that the NLO corrections in the the FFN
prescription have been calculated \cite{LRSN}, and that the leading
(collinear) logarithms of the type $\alpha_s^i\log^i(Q^2/M_H^2)$ have
been extracted in analytic form \cite{buza}. Thus the S-ACOT
subtraction term can be constructed as well.

Finally, we emphasize that the choice of a prescription for dealing
with quark masses in the hard scattering coefficients for deeply
inelastic scattering is a separate issue from the choice of definition
of the parton distribution functions. For all of the prescriptions
discussed here, one uses the standard \MSbar\ definition of parton
distributions.

\small{\section*{Acknowledgments}
We thank J.C.~Collins, R.J.~Scalise, R.S.~Thorne, and W.-K.~Tung for
valuable discussions and we thank S.~Kretzer for making his NLO code
available to us. We thank the CERN Theory Division, Fermilab Theory
Group, and the University of Oregon for their kind hospitality during
the period in which part of this research was carried out.  This work
is supported by the U.S.  Department of Energy, the Lightner-Sams
Foundation, and in part by the EU TMR contract FMRX-CT98-0194 (DG 12 -
MIHT).}



\end{document}